\documentclass[prb,twocolumn,amsmath,amsfonts,8pt,floatfix]{revtex4-1}
\usepackage[utf8]{inputenc}
\usepackage{amsmath}
\usepackage{amsfonts}
\usepackage{amssymb}
\pdfoutput=1
\usepackage{amsmath}
\usepackage{amssymb}
\usepackage{graphicx}
\usepackage{epstopdf}
\graphicspath{{figures/}}
\usepackage{color}   
\usepackage[ocgcolorlinks]{hyperref}
    \usepackage{dcolumn}
   \newcolumntype{d}[1]{D{.}{.}{#1}}
\hypersetup{
    colorlinks=true, 
    linktoc=all,     
    colorlinks=true,       
    linkcolor=red,          
    citecolor=green,        
    filecolor=magenta,      
    urlcolor=blue           
    plainpages=false,
    pdfpagelabels
}
\usepackage{float}
    \usepackage{xstring}
    \usepackage{multirow}
\usepackage{float}
\usepackage{etoolbox}
\usepackage{pgffor}
\usepackage{forarray}
\usepackage{pgffor}%
    \usepackage{dcolumn}
   \newcolumntype{d}[1]{D{.}{.}{#1}}

\pdfoutput=1
\newcommand\setprefix[2]{\expandafter\def\csname#1\endcsname{#2}}
\newcommand\getprefix[1]{\csname#1\endcsname}
\makeatletter
\g@addto@macro\bfseries{\boldmath}
\makeatother

\newcommand{\up}{{\uparrow}}
\newcommand{\down}{{\downarrow}}

\begin{document}
\title{Concept of chemical bond and aromaticity based on quantum information theory}
\author{T. Szilv\'asi$^1$, G. Barcza$^2$, \"O. Legeza$^2$}
\affiliation{$^1$Department of Inorganic and Analytical Chemistry, Budapest University of Technology and Economics, H-1111 Budapest, Hungary\\
$^2$MTA-WRCP Strongly Correlated Systems "Lend\"ulet" Research group, H-1525 Budapest, Hungary}

\begin{abstract}
Quantum information theory (QIT) emerged in physics as standard technique
to extract relevant information from quantum systems. It has
already contributed to the development of novel fields like 
quantum computing, quantum cryptography, and quantum complexity.
This arises the question what
information is stored according to QIT in molecules which are 
inherently quantum systems as well. 
Rigorous analysis of the central quantities of QIT
on systematic
series of molecules offered the introduction of the concept of chemical
bond and aromaticity directly from physical principles and notions.
We identify covalent bond, donor-acceptor dative bond,
multiple bond, charge-shift bond, and aromaticity 
indicating unified picture
of fundamental chemical models from ab initio.
\end{abstract}


\date{\today}
\maketitle


Extension of information theory\cite{Haken} for quantum systems,
called quantum information
theory (QIT)\cite{Nielsen-2000,Preskill,Wilde-2013}, leads to emergence of
non classical correlations, entanglement.~\cite{Horodecki-2009}
In the past decade various concepts of 
QIT have matured to widely used tools in quantum many body physics.~\cite{Amico-2008}
Several well known quantities have been redefined
in term of entanglement shedding new light in our understanding of 
quantum systems.

In chemistry such concepts have appeared only recently
\cite{Legeza-2003c,Pipek-2009,McKemmish-2011,Boguslawski-2015} although 
correlations among components of a 
finite system, like orbitals,
can be regarded as exchanging information among such parties.
For example, single orbital entropy 
provides information about how much an orbital is entangled with the
rest of the system.~\cite{Legeza-2003c}
In addition, two-orbital mutual information\cite{Legeza-2006,Rissler-2006}
determines how orbitals communicate with each other, i.e., 
it gives the correlation between two orbitals
as they are embedded in the whole system. 
The mutual information, however, contains correlations of both classical
and quantum origin.~\cite{Modi-2010}
Such central quantities to describe orbital correlations
have already been applied recently by some of us
to analyze copper-oxide clusters\cite{Barcza-2011},
to dissect electron correlation effects in bond-forming and bond-breaking
processes.~\cite{Boguslawski-2013a,Mottet-2014,Duperrouzel-2015}
In addition, entanglement structures have also been determined in
photosystem II~\cite{Kurashige-2013} and 
orbital entanglement analysis of the
Ru–NO bond in a Ruthenium nitrosyl complex has also been
carried out.~\cite{Freitag-2015} 
In all these works, correlations among the orbitals were measured in
terms of quantum information entropies which are
weighted averages of the eigenvalue spectrum of the various subsystem
density matrices (see Eq.(\ref{eq:si})).
Therefore, the more detailed information encoded in the 
the eigenvalue spectrum~\cite{Peschel-1999,Li-2008}
and in the structure of the corresponding eigenstates
of 
reduced density matrix has been lost.

In this work, we present a systematic analysis of 
(multi-)orbital entanglement together with the 
probability distribution of eigenstates given by
the corresponding reduced density matrices
for all possible realizations of the two-orbital subsystems
in series of handful of molecules. 
As a result, we find strong connection between our approach and
basic chemical models which allow us to  
describe covalent bonds, donor-acceptor dative bonds, 
multiple bonds, charge-shift bond, and aromaticity from a unified point of view.


{\sl Theoretical background:}
When a system is split into two parts called a bipartite system
(the two parts often called Alice and Bob), the Hilbert space
is ${\cal H}={\cal H}^{(\rm A)}\otimes{\cal H}^{(\rm B)}$.
If the system can be described by a pure state,
the wave function is a linear combination of the tensor product of the
basis functions of the two subsystems
$(|\phi^{({\rm A})}_\alpha\rangle, |\phi^{({\rm B})}_\beta\rangle)$, i.e,
$|\Psi\rangle=\sum_{\alpha\beta}C_{\alpha\beta}|\phi^{({\rm A})}_\alpha\rangle\otimes|\phi^{({\rm B})}_\beta\rangle$,
where $C_{\alpha\beta}$ is a complex matrix in general.
The correlations between the two subsystems
is fully quantum mechanical and
called entanglement.
The wave function can also be written as a single sum
of the product of transformed basis states due to
Schmidt decomposition,
i.e.,
$|\Psi\rangle = \sum_{\alpha=1}^r \sqrt \omega_\alpha |\xi^{({\rm A})}_\alpha
\rangle\otimes|\xi^{({\rm B})}_\alpha\rangle$,
where $r\leq\min(\dim {\cal H}^{(\rm A)},\dim {\cal H}^{(\rm B)})$, 
$\omega_\alpha \geq 0$, and $\sum_{\alpha=1}^r  \omega_\alpha=1$.
The square of the $\sqrt \omega_\alpha$ Schmidt values are also equal to
the eigenvalues of the so-called reduced density matrix,
$\rho^{(A)}$, formed by tracing out one subsystem\cite{Nielsen-2000}, i.e.,
$\rho^{(A)}={\rm Tr_B}|\Psi\rangle\langle\Psi|$, thus
$\rho^{(\rm A)}_{\alpha,\alpha^\prime}=\sum_\beta C_{\alpha\beta} C_{\alpha^\prime \beta}^*$.
$|\xi^{({\rm A})}_\alpha\rangle$ and $|\xi^{({\rm B})}_\alpha\rangle$ are the
eigenstates of $\rho^{(\rm A)}$ and $\rho^{(\rm B)}$, respectively. 
If $r=1$ the wave function $|\Psi\rangle$ is a product state and the system is
called separable,
otherwise it is said to be entangled.

In general,
both subsystems are in a mixed state and the total system cannot be
written as a single product of the states of the two subsystems.
There are various quantities introduced
to measure the mixedness of the subsystems and the strength of
entanglement \cite{Amico-2008} but they all
must fulfill an important criterion namely entanglement
monotonicity \cite{Vidal-2000,Horodecki-2001,Szalay-2013}.
This means that the quantity cannot increase by local operations and classical
communication (LOCC).
A widely used quantity is the entanglement entropy
given by the von-Neumann entropy of the reduced density matrix calculated as
\begin{equation}
S(\rho^{(\rm A)})=-\sum_{\alpha} \omega_{\alpha}^{(\rm A)}\ln \omega_{\alpha}^{(\rm A)}.
\label{eq:si}
\end{equation}
There are various possibilities to split a system into two or
several subsystems.
If subsystem $(\rm A)$ contains a single orbital and subsystem $(\rm B)$ 
the rest of the orbitals,
the single-orbital entropy, $S(\rho^{(i)})$,
can be calculated with $i=1,\ldots N$,
where $N$ is the number of orbitals.~\cite{Legeza-2003c}
Assuming four basis states per orbital, i.e.,
$|\phi^{(i)}_\alpha\rangle\in\{
|0\rangle, |\down\rangle, |\up\rangle, |\up\down \rangle \}$,
the theoretical maximum of $S(\rho^{(i)})$ is $\ln 4$
with $\omega^{(i)}_{\alpha}=1/4$
for all $\alpha=1,\ldots,4$.

The correlation between two orbitals, $i$
and $j$, as they are embedded in the whole system
is given by the two-orbital mutual 
information\cite{Legeza-2006,Rissler-2006}
\begin{equation}
I^{(i,j)} = S(\rho^{(i)}) + S(\rho^{(j)}) - S(\rho^{(i,j)}) \,,
\label{eq:mut}
\end{equation}
where $S(\rho^{(i)})$, $S(\rho^{(j)})$, $S(\rho^{(i,j)})$ are the
single-orbital and two-orbital
entropies, respectively.
In this case
$\rho^{(i,j)}$ is also a mixed state thus the
mutual information contains correlations of both classical
and quantum origin.~\cite{Modi-2010}
An eigenvalue of $\rho^{(i,j)}$ denoted by
$\omega^{(i,j)}_\alpha$ corresponds to eigenvector
$|\xi^{(i,j)}_\alpha\rangle
=\sum_{\alpha_i,\alpha_j} C^{(i,j)}_{\alpha_i,\alpha_j}(\alpha)
|\phi_{\alpha_i}\rangle\otimes|\phi_{\alpha_j}\rangle$
where $\alpha_i$ and $\alpha_j$ run from 1 to 4 and $\alpha=1,\ldots 16$.
Again the theoretical maximum of $I^{(i,j)}=\ln 16$ corresponds to
maximally entangled pure two-orbital state with
$\omega^{(i,j)}=[1,0\ldots,0]$.
This also means that orbital $i$ and $j$ are
in maximally mixed state with
$\omega^{(i)}_\alpha=\omega^{(j)}_\alpha=1/4$ for all $\alpha$
and $S(\rho^{(i)})=S(\rho^{(j)})=\ln 4$,
but orbital pair state $(i,j)$ is in a pure state with $S(\rho^{(i,j)})=0$.

In case of quantum chemical systems finite number of electrons
are correlated on finite number of orbitals and the
number of electrons, $n$, as well as the total spin projection, $s^z$,
are good quantum numbers. 
$\rho^{(i,j)}$ also commute with the Hamilton operator and it
has a block diagonal
structure ~\cite{Boguslawski-2013a,Barcza-2014,Szalay-2015} and
the $\alpha^{\rm th}$ eigenstate of the two-orbital subsystem
with quantum number pair $(n,s^z)$
(with $n=0,\ldots,4$ and $s^z\in\{-1,-1/2,0,1/2,1$\}
can be written as a linear combinations of basis states in the
corresponding subspace.
The set of $c^{(i,j)}_{\alpha_i,\alpha_j}$
coefficients corresponding to basis states with quantum number pair $(n,s^z)$
will be labeled by a vector $c^{(i,j)}_\alpha(n,s^z)$
in order to use a compact notation for the rest of the paper.
Therefore,
$c^{(i,j)}_\alpha(0,0)$ corresponds to $|0,0\rangle$,
$c^{(i,j)}_\alpha(1,-\frac{1}{2})$ to
$\{|0,\down\rangle,|\down,0\rangle\}$,
$c^{(i,j)}_\alpha(1,\frac{1}{2})$ to
$\{|0,\up\rangle,|\up,0\rangle\}$,
$c^{(i,j)}_\alpha(2,0)$ to
$\{|0, \up\down \rangle$, $|\up, \down \rangle$, $|\down, \up  \rangle$,
$|\up\down, 0 \rangle\}$,
$c^{(i,j)}_\alpha(2,-1)$ to
$|\down,\down \rangle$,
$c^{(i,j)}_\alpha(2,1)$ to
$|\up, \up\rangle$,
$c^{(i,j)}_\alpha(3,-\frac{1}{2})$ to
$\{|\up\down, \down \rangle$, $|\down, \up\down \rangle\}$,
$c^{(i,j)}_\alpha(3,\frac{1}{2})$ to
$|\up\down, \up \rangle$, $|\up, \up\down \rangle$\},
and
$c^{(i,j)}_\alpha(4,0)$ to $|\up\down,\up\down\rangle$.
In the rest of the paper, $\omega^{(i,j)}_\alpha$ eigenvalues will be
ordered decreasingly, i.e, the largest value will
correspond to $\alpha=1$ and the smallest to $\alpha=16$.
We will use the term relevant eigenvalue if it is one or two orders
of magnitude larger than the remaining ones.
The elements of the $c^{(i,j)}_\alpha(n,s^z)$ vector
will be given in terms of its largest element in order to show the
ratio among the coefficients.
Furthermore, in some cases the eqality among the various
$c^{(i,j)}_\alpha(n,s^z)$ vectors discussed in the next section will hold
up to a spin reversal.
\begin{figure*}
\centerline{
\includegraphics[scale=1]{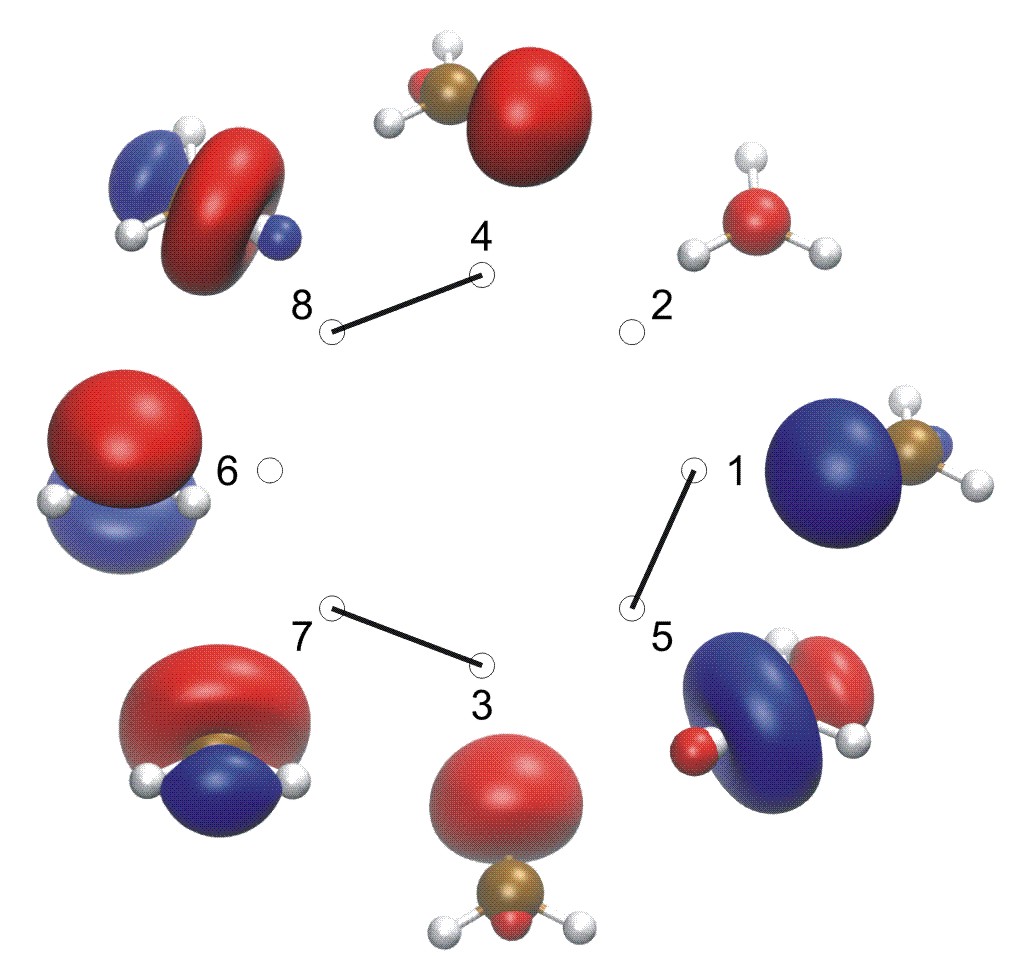}
\hskip 0.25cm
\includegraphics[scale=1]{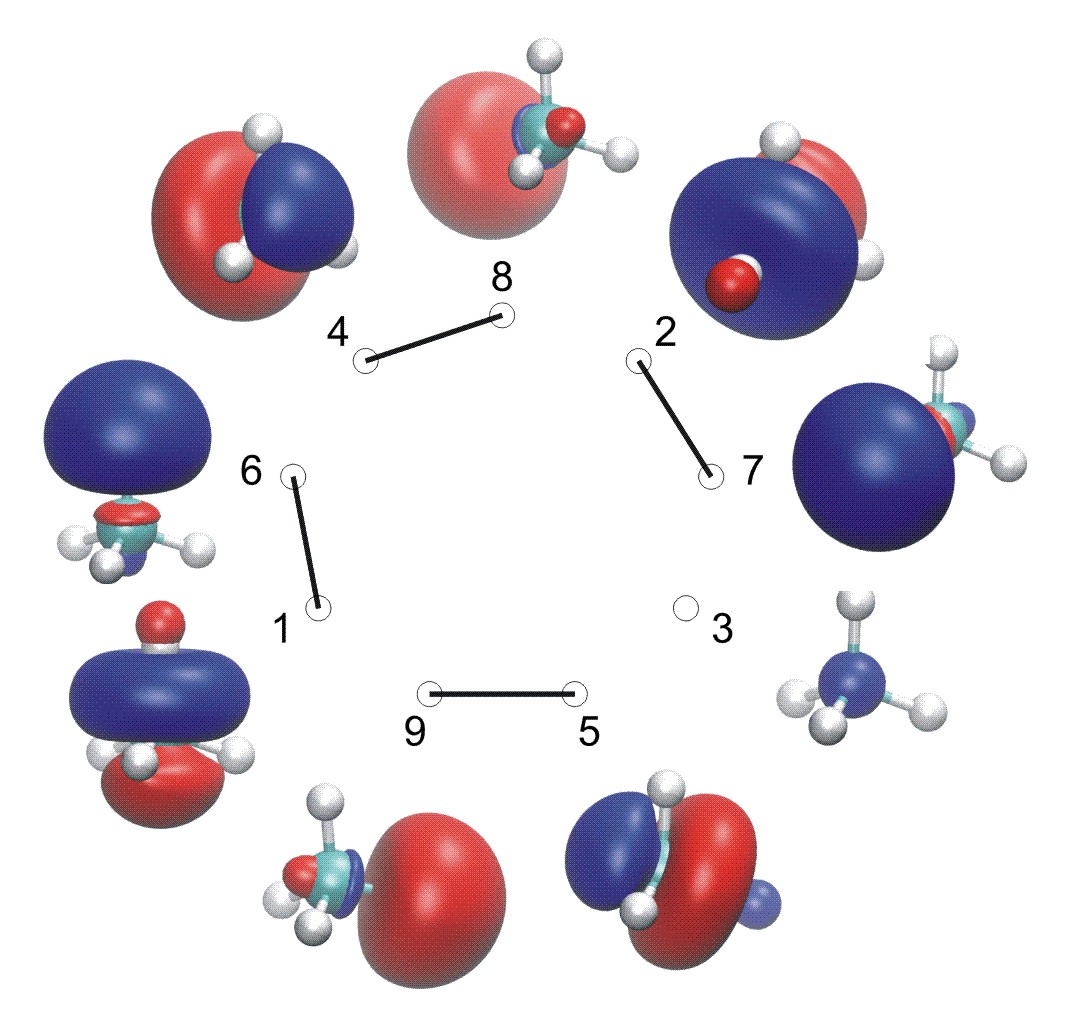}
\hskip 0.25cm
\includegraphics[scale=1]{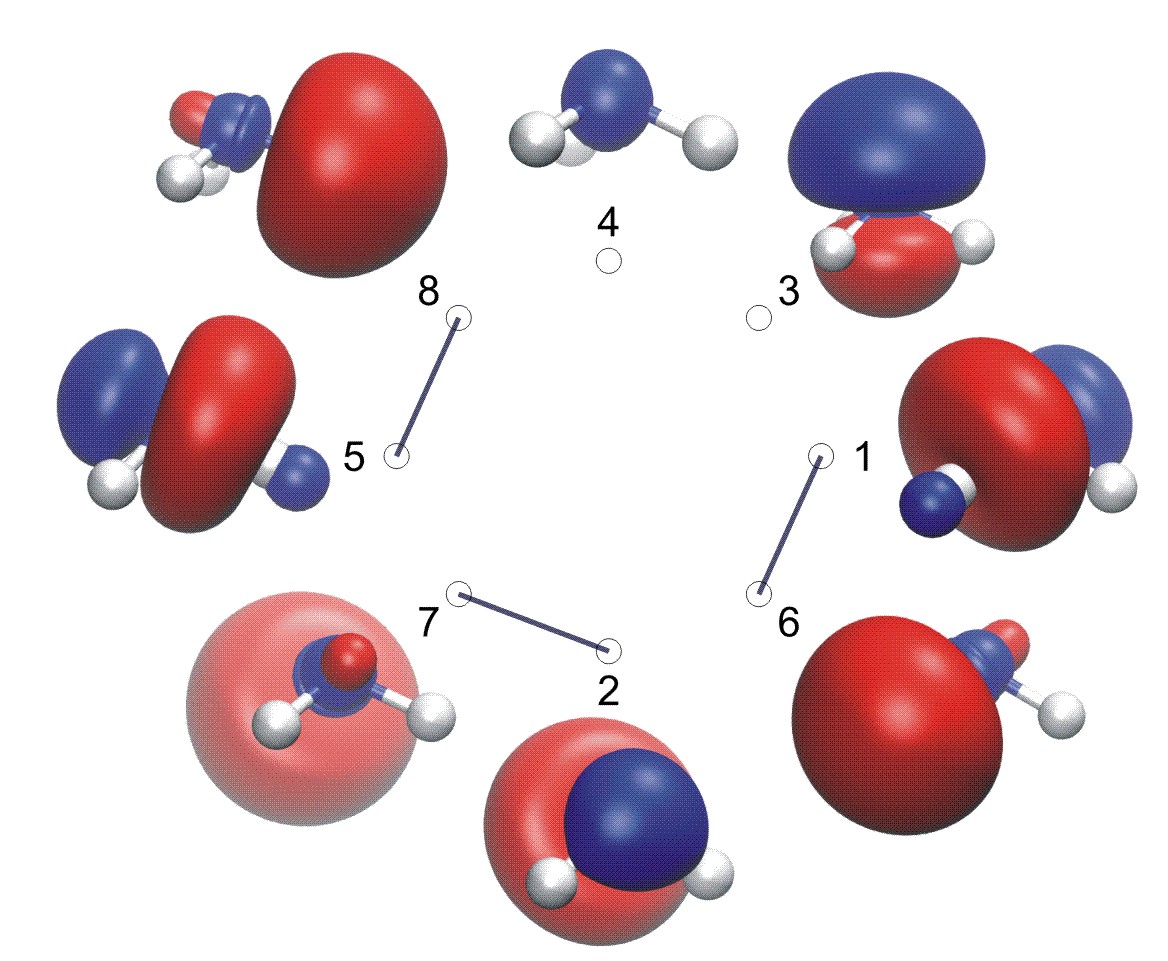}
}
\vskip -4.3cm
\centerline{\hskip -5.2cm a)\hskip 4.8cm b) \hskip 4.8cm c)}
\vskip 4.3cm
\vskip 0.25cm
\centerline{
\includegraphics[scale=1]{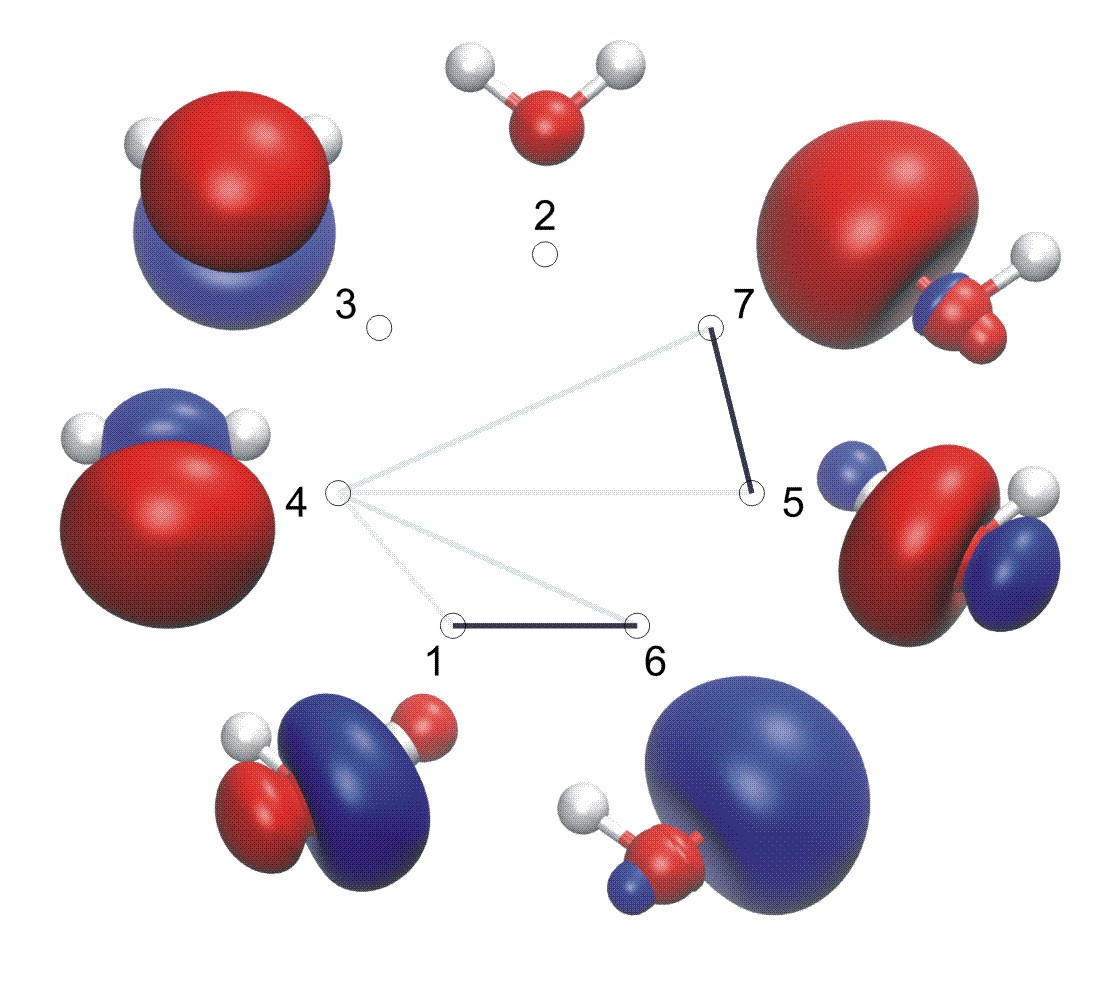}
\hskip 0.25cm
\includegraphics[scale=1]{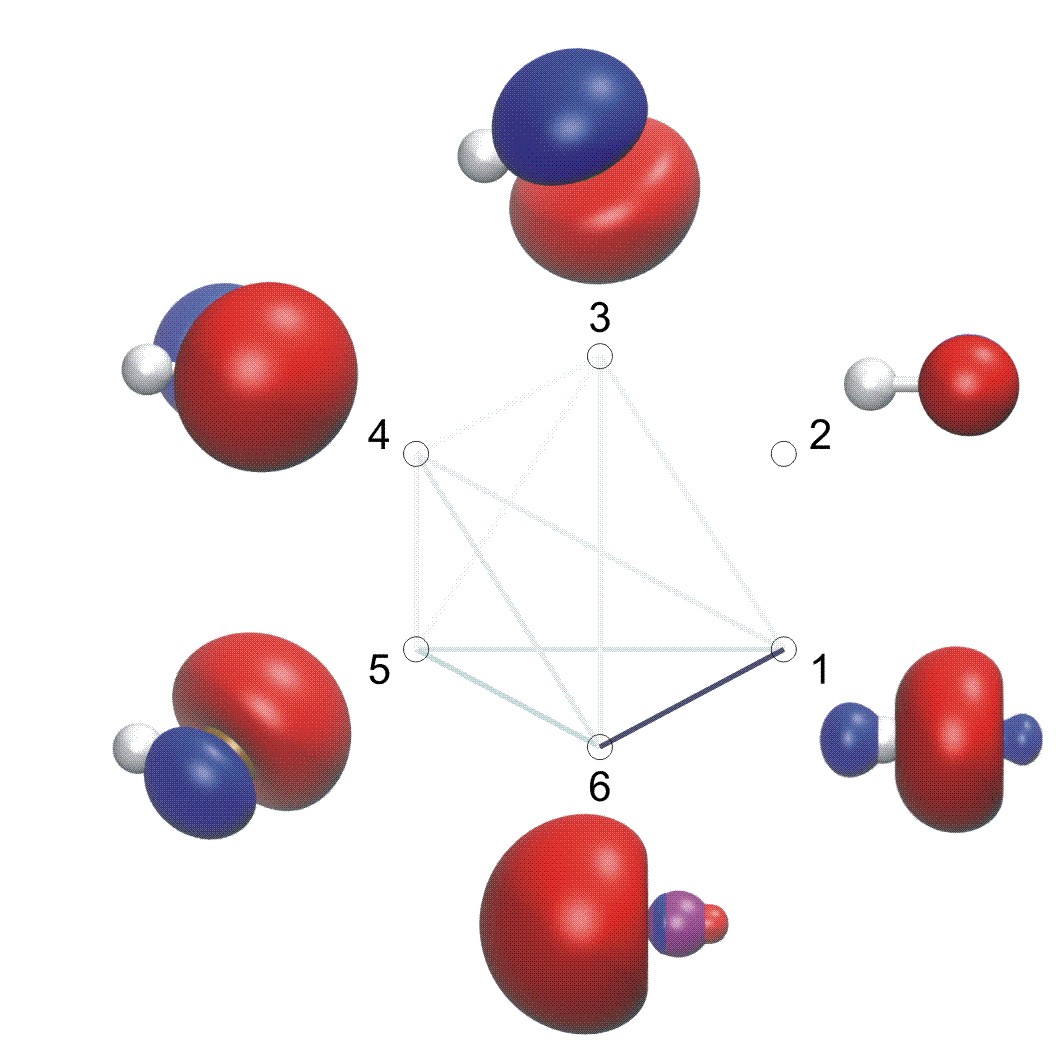}
\hskip 0.25cm
\includegraphics[scale=1]{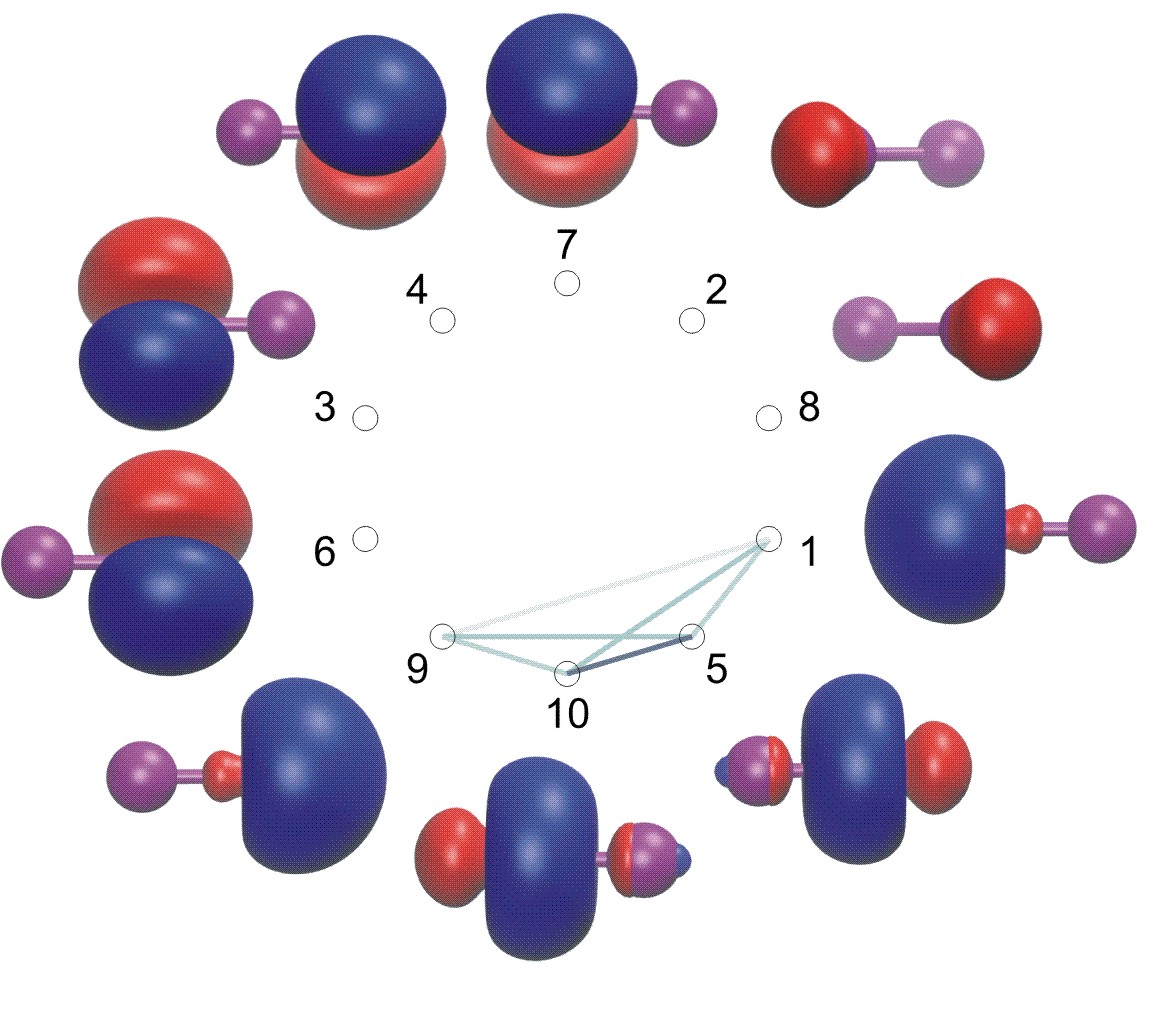}
}
\vskip -4.3cm
\centerline{\hskip -5.2cm d)\hskip 4.8cm e) \hskip 4.8cm f)}
\vskip 4.3cm
\vskip 0.25cm
\centerline{
\includegraphics[scale=1]{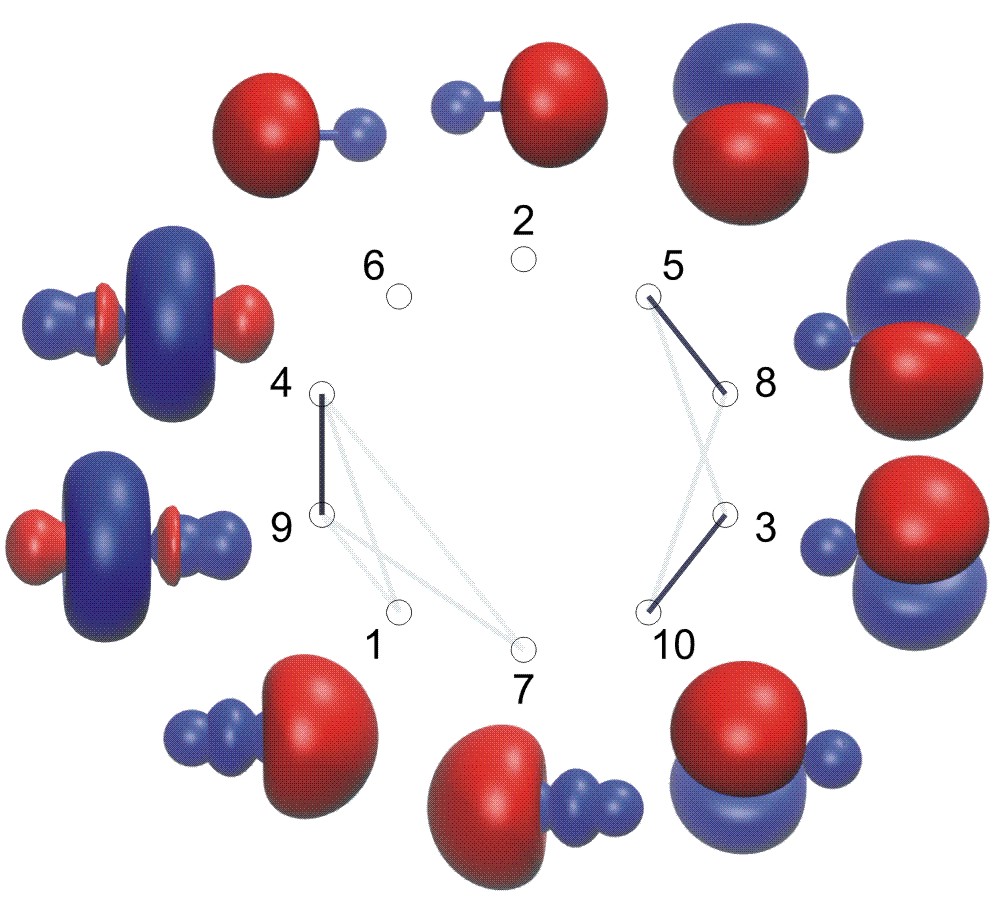}
\hskip 0.25cm
\includegraphics[scale=1]{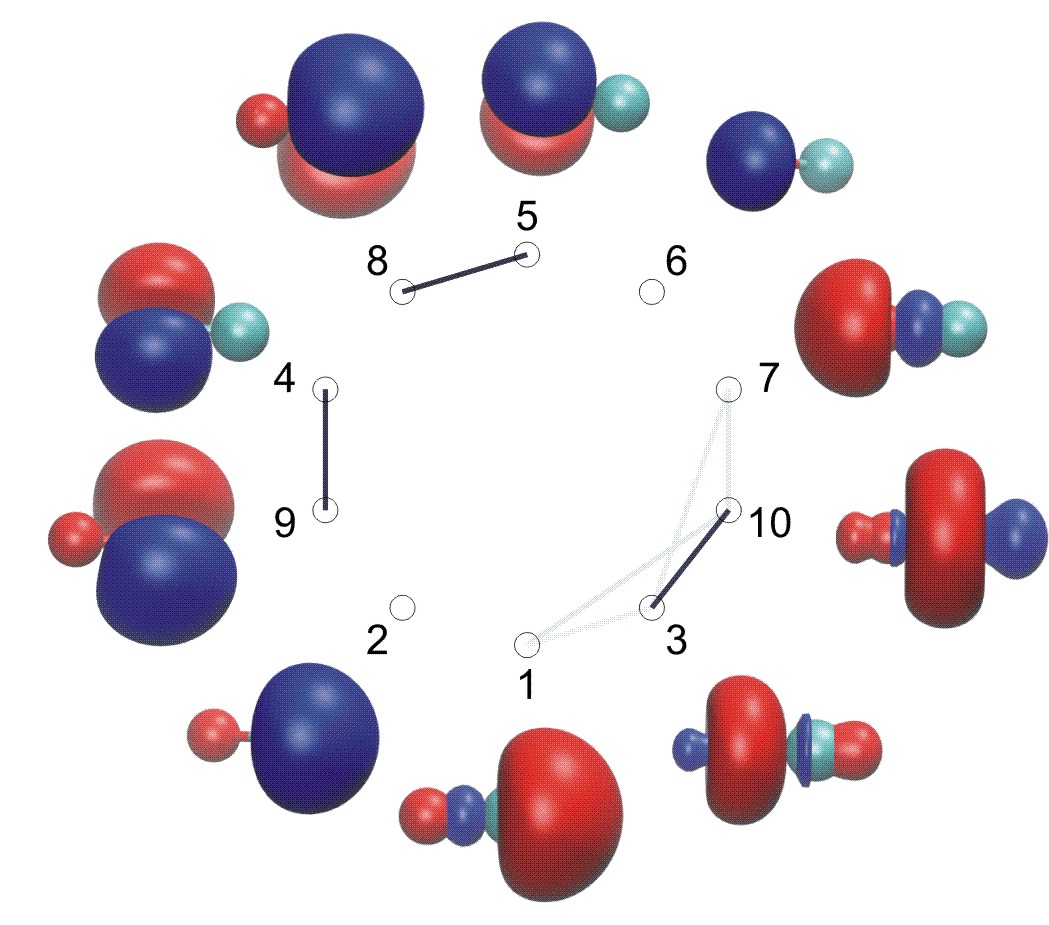}
\hskip 0.25cm
\hskip 1cm 
\includegraphics[scale=0.5]{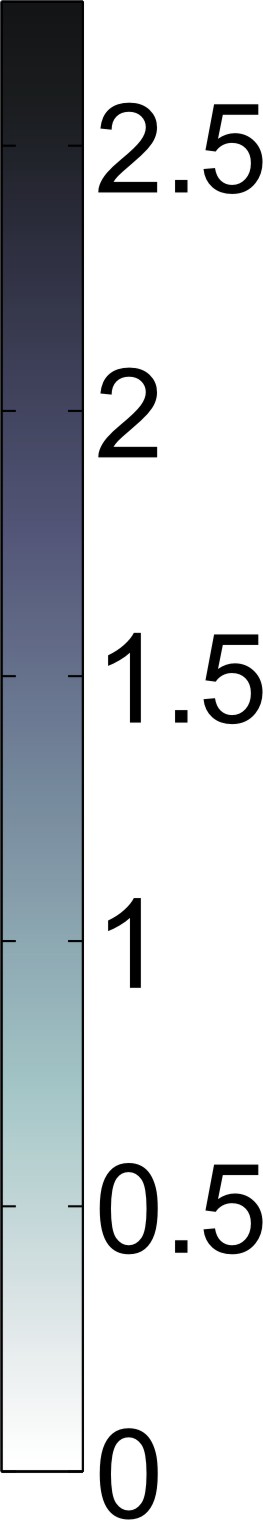}
\hskip 3cm
}
\vskip -4.3cm
\centerline{\hskip -5.2cm g)\hskip 4.8cm h) \hskip 4.8cm i)}
\vskip 4.3cm
\caption{Graphical representation of mutual information for basic small molecules: BH$_3$ (a), CH$_4$ (b), NH$_3$ (c), H$_2$O (d), HF (e), F$_2$ (f), N$_2$ (g), and CO (h). The color of the bonds 
according to the color scale (i)
indicates the mutual information of the connected two sites.
The shape of the molecular orbitals (isovalue 0.05) are shown directly near the sites. The blue and red colors of the orbitals refer to the sign of the wave function. Atoms are labeled with their usual colors; H, B, C, N, O, and F atoms marked with white, brown, tan, blue, red, and purple colors, respectively.
}
\label{fig:small}
\end{figure*}
%
%
%
\vskip 0.1cm
{\bf Results and Discussion}\\
Fig.~\ref{fig:small} shows the mutual information results of eight small
molecules, BH$_3$, CH$_4$, NH$_3$, H$_2$O, HF, F$_2$, N$_2$, and CO.
Numbers on each pictures only refer to different sites which orbitals are
also shown around. Mutual information is indicated by the
color of the lines between the sites. The lack of strong visible line
between sites is the sign of negligible communication; these orbital
correlations do not contain relevant information. The maximum of the
mutual information is marked by the theoretical limit, $\ln 16\simeq 2.77$,
as it was discussed earlier.

The mutual information analysis of BH$_3$ reveals three strong
entanglements given by $I^{(1,5)}=I^{(3,7)}=I^{(4,8)}=2.68$ (Fig.~\ref{fig:small}a).
All other 25 possible connections are zero within numerical accuracy.
Therefore, site \#2 and \#6 do not communicate with other sites which is in
accordance with the chemical insight as site \#2 is the inert 1s core
orbital while site \#6 is the empty 2p orbital of the boron.
The calculated mutual information of the correlated sites are identical and
very close to the theoretical upper limit as their two-orbital reduced
density matrix is dominated by a single eigenvalue
$\omega^{(1,5)}_1 = \omega^{(3,7)}_1 = \omega^{(4,8)}_1\simeq 0.992$. 
In the corresponding eigenvector, elements 
$|0, \up\down \rangle$, $|\up, \down \rangle$, $|\down, \up  \rangle$,
$|\up\down, 0 \rangle$,
have almost the same relevance in the description of the
interaction, i.e., $c^{(1,5)}_1(2,0) = c^{(3,7)}_1(2,0) = c^{(4,8)}_1(2,0) = 0.541\times[0.80, -1, 1, 0.88]$.

Similar picture is found for CH$_4$ molecule. Core orbital, site \#3, does
not entangle with other sites (Fig.~\ref{fig:small}b) while there are four pairs of sites with considerable mutual information
$I^{(1,6)}=I^{(2,7)}=I^{(4,8)}=I^{(5,9)}=2.68$.
Orbital pictures on Fig.~\ref{fig:small}b show that in
all four cases one hydrogen and carbon orbitals present strong mutual
information.
The corresponding relevant eigenvector elements are similar to the results of BH$_3$
$c^{(1,6)}_1(2,0)=c^{(2,7)}_1(2,0)=c^{(4,8)}_1(2,0)=c^{(5,9)}_1(2,0)=0.543\times [0.79, -1, 1,  0.87]$.
Mutual information analysis of other small molecules like NH$_3$, H$_2$O,
HF, and F$_2$ (Fig.~\ref{fig:small}c-f) shows similar overall picture to BH$_3$ and
CH$_4$ indicating 3, 2, 1, and 1 very strong mutual information
between sites, respectively (see also in SI \ref{sec:covalent}). 
These correlations always connect orbitals
on different atoms and thus can be assigned to chemical bonds. The
calculated mutual information of NH$_3$, H$_2$O, HF, and F$_2$ is,
however, only 2.52, 2.23, 1.85, and 1.47 smaller for these correlations
than that in the case of BH$_3$ and CH$_4$.
Deviation from the theoretical limit reflects the relative difference of
the elements of the eigenvector and the presence of additional correlations of sites.
For NH$_3$, one eigenvalue dominates
$\omega^{(1,6)}_1$ = $\omega^{(2,7)}_1$ = $\omega^{(5,8)}_1$ = 0.967 with moderate difference in the weight of the eigenvector elements,
$c^{(1,6)}_1(2,0)=c^{(2,7)}_1(2,0)=c^{(5,8)}_1(2,0)= 0.541\times[0.73, -1,  1,   0.94]$.

For H$_2$O, the eigenvector elements have even more asymmetric distribution compared to NH$_3$
$\left(c^{(1,6)}_1(2,0)= c^{(5,7)}_1(2,0)=0.539\times[0.69, -1, 1, 0.98]\right)$. Besides,
site \#4, the lone pair in the plane, has also
minor contribution to the mutual information picture,
$I^{(4,6)}$ = $I^{(4,7)}$ = 0.30 and $I^{(4,1)}$ = $I^{(4,5)}$ = 0.22. These values are one order of magnitude smaller than the usual ones
$I^{(1,6)}=I^{(5,7)}=2.23$.
Further analysis reveals that this secondary effect comes from
the three electron regime 
given by
the second largest eigenvalue,
$\omega_2^{(4,6)}=0.249$
and the corresponding
vector elements of $c^{(4,6)}(3, \pm \frac{1}{2})$. These secondary
correlations are present between a site with two electrons and the two
sites with large mutual information which may be regarded as
hyperconjugative effect between a lone pair and an adjacent bond.
Similar effects are observed for HF and F$_2$ as well which is also partly responsible 
for the lower value of $I$ which is assigned to the chemical bond. 
However, the main reason is the large value of the second largest eigenvalue.
For HF, $\omega^{(1,6)}_2$ =$\omega^{(1,6)}_3$ = 0.118, while for F$_2$
$\omega^{(5,10)}_2$ =$\omega^{(5,10)}_3$ = 0.174. It turns out that this second and third 
eigenvalues are associated with the three-electron regime. For HF,
$c^{(1,6)}_2(3,\pm\frac{1}{2})=c^{(1,6)}_3(3,\pm\frac{1}{2})= 0.882\times[1, -0.53]$, for F$_2$
$c^{(5,10)}_2(3,\pm\frac{1}{2})=c^{(5,10)}_3(3,\pm\frac{1}{2})= 0.707\times[1, 1]$.
It formally means that an extra electron resonates between the two sites which is consistent 
with picture of charge-shift bond.\cite{Hiberty-1992}\cite{Hiberty-2009}
The orbital pictures of Fig.~\ref{fig:small} underline this correspondence.
\begin{figure*}
\centerline{
\includegraphics[scale=1.1]{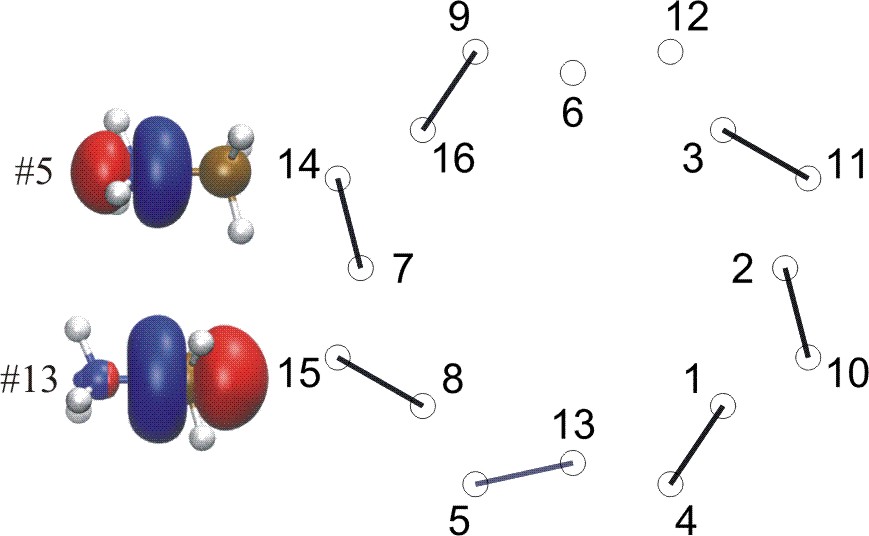}
\hskip 0.3cm
\includegraphics[scale=1.1]{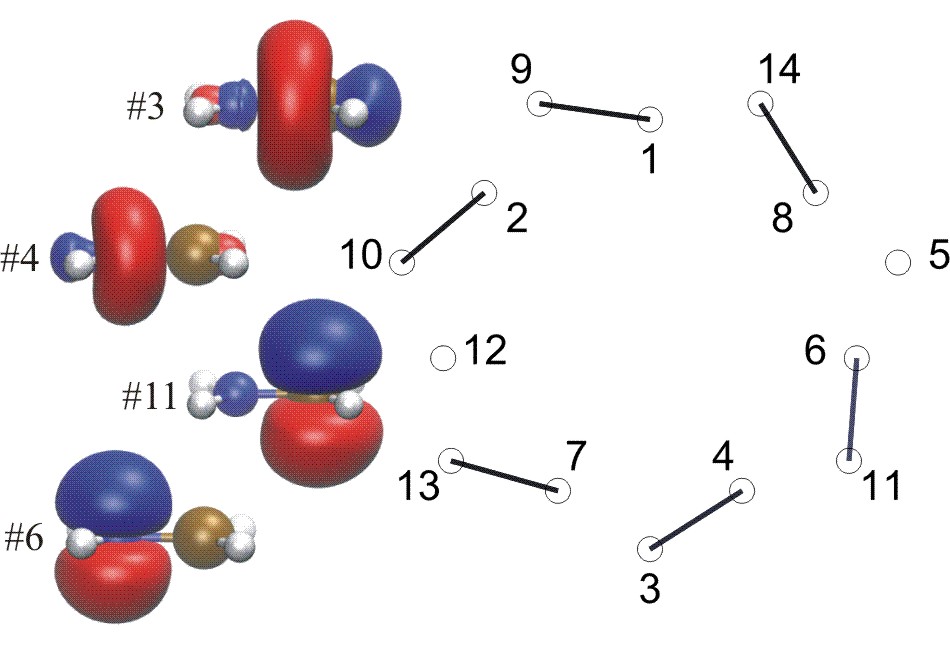}
\hskip 0.3cm
\includegraphics[scale=1.1]{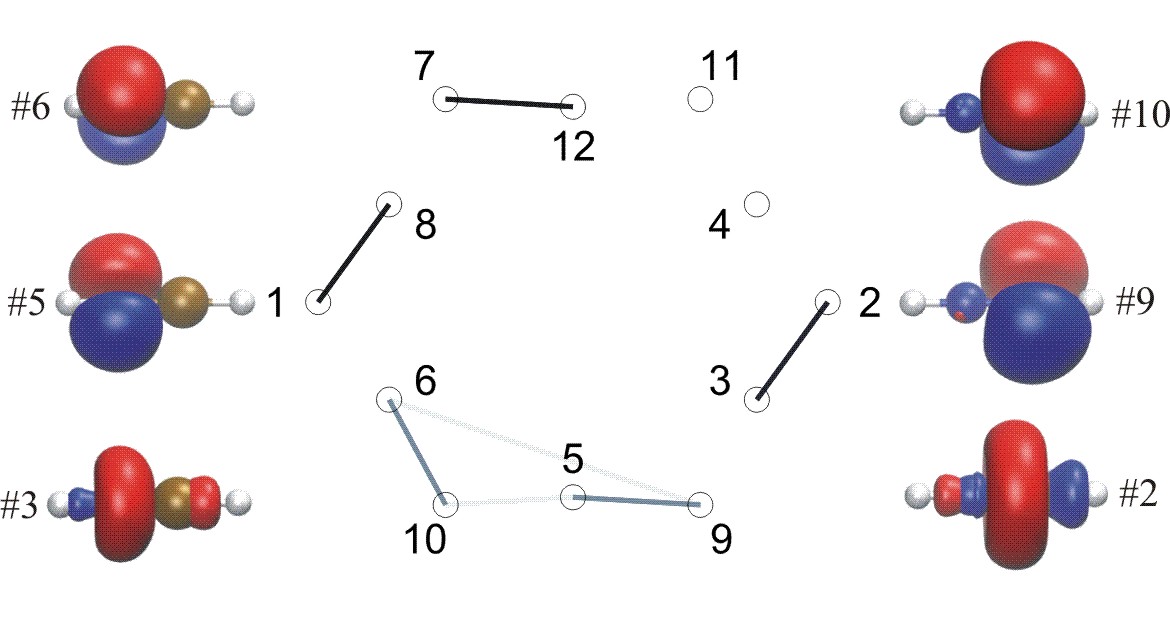}
}
\vskip -3.4cm
\centerline{\hskip -6.1cm a)\hskip 4.6cm b) \hskip 4.8cm c)}
\vskip 3.4cm
\vskip 0.2cm
\centerline{
\includegraphics[scale=1.1]{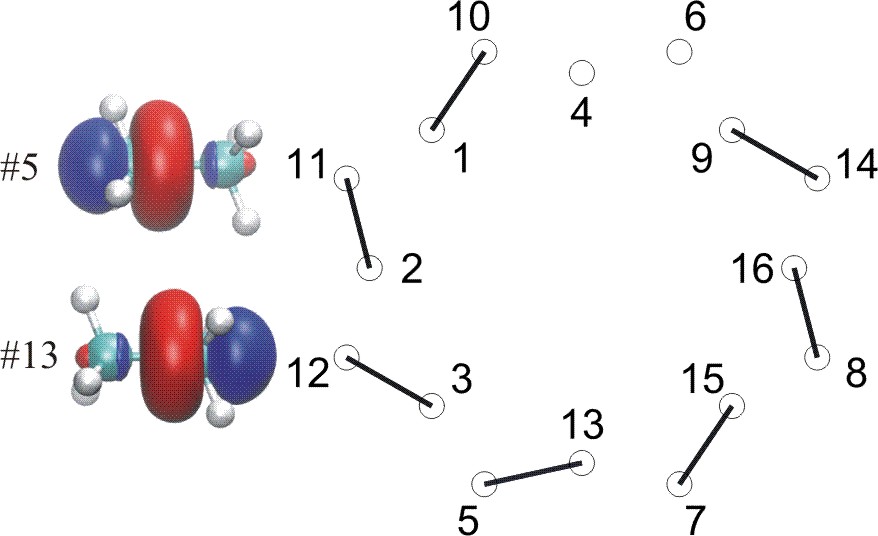}
\hskip 0.3cm
\includegraphics[scale=1.1]{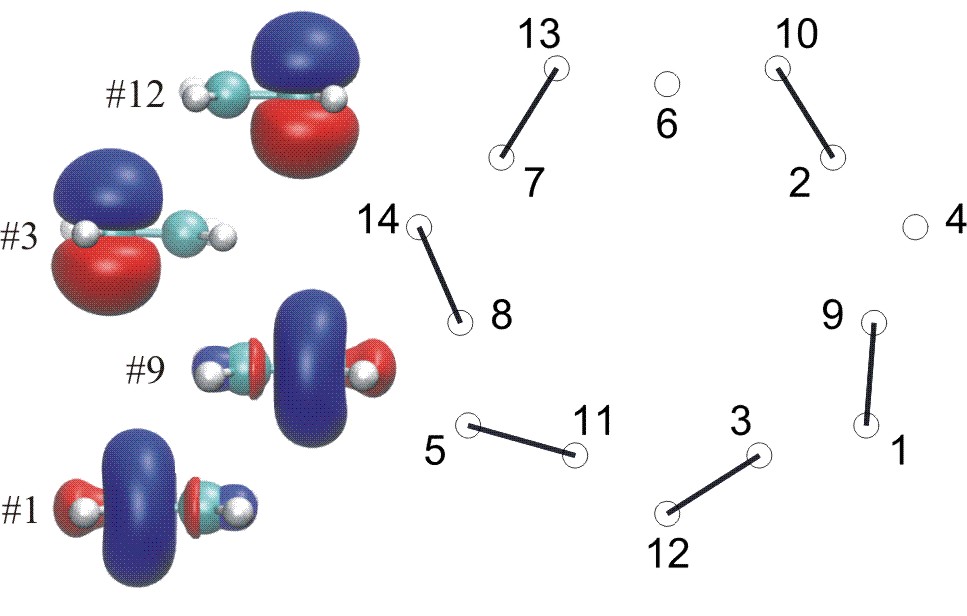}
\hskip 0.3cm
\includegraphics[scale=1.1]{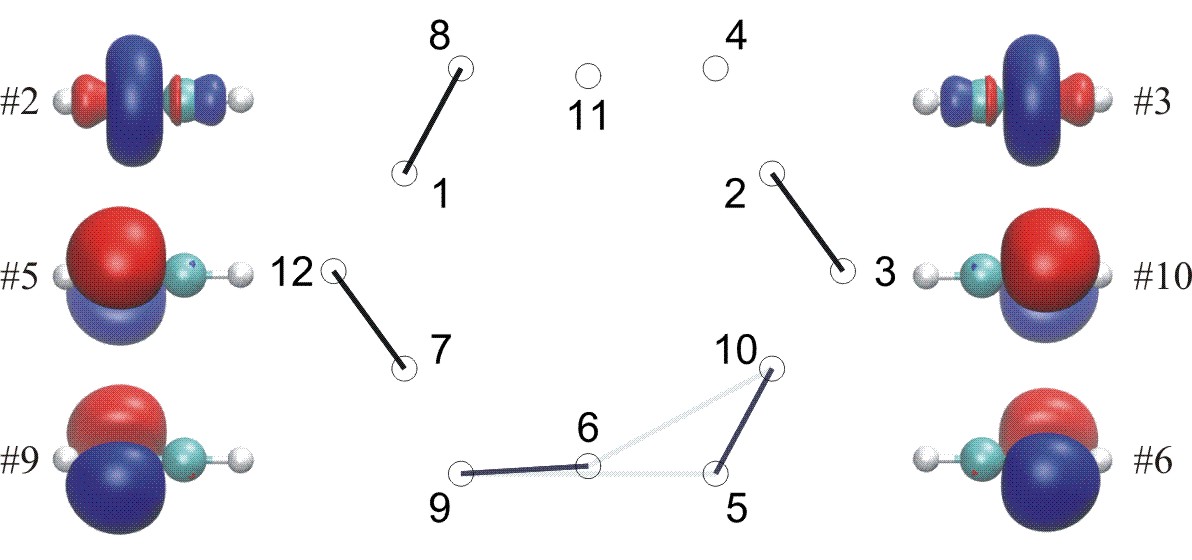}
}
\vskip -3.4cm
\centerline{\hskip -6.1cm d)\hskip 4.6cm e) \hskip 4.8cm f)}
\vskip 3.4cm
\vskip 0.2cm
\centerline{
\includegraphics[scale=1.1]{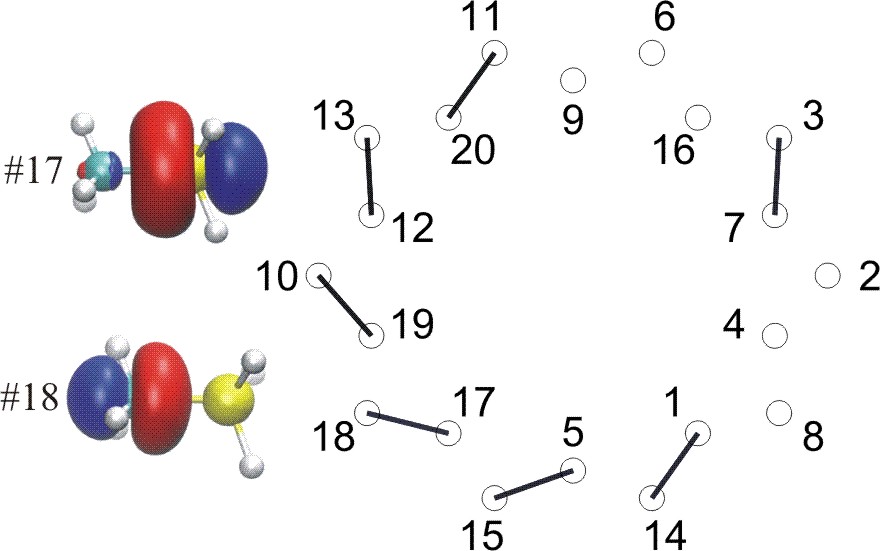}
\hskip 0.3cm
\includegraphics[scale=1.1]{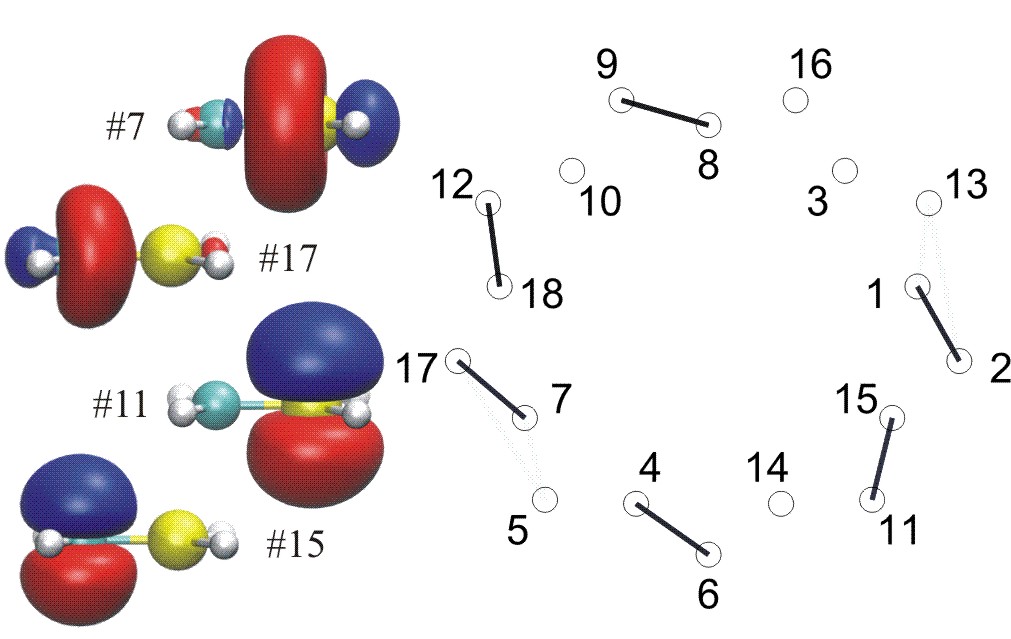}
\hskip 0.3cm
\includegraphics[scale=1.1]{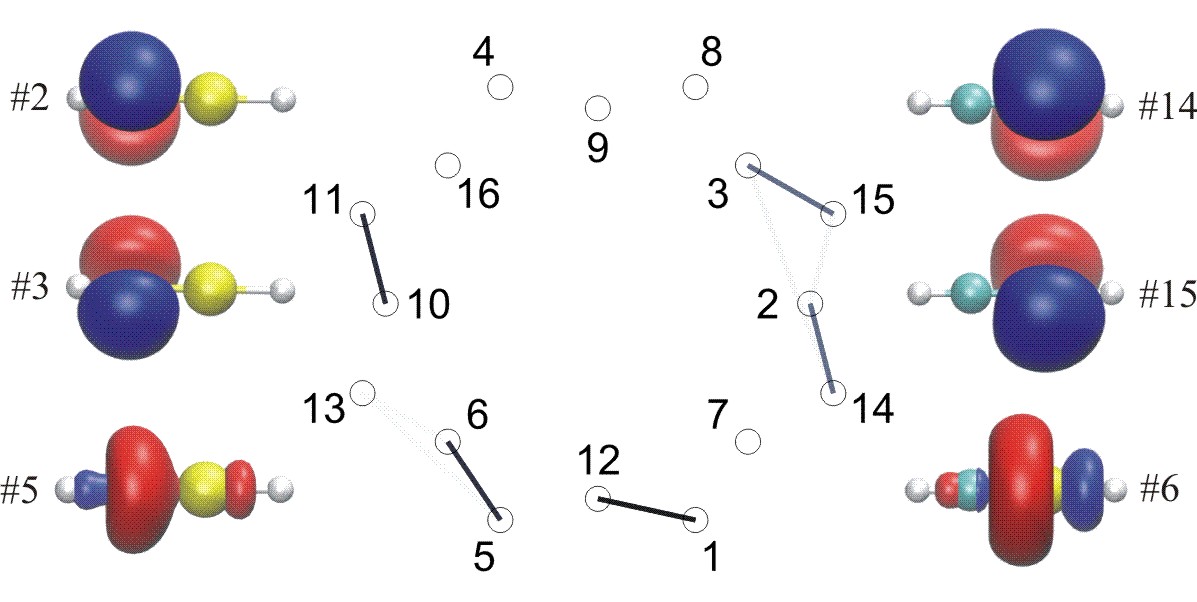}
}
\vskip -3.4cm
\centerline{\hskip -6.1cm g)\hskip 4.6cm h) \hskip 4.8cm i)}
\vskip 3.4cm
\caption{Graphical representation of mutual information for the series of NH$_3$BH$_3$ (a), NH$_2$BH$_2$ (b), NHBH (c), CH$_3$CH$_3$ (d), CH$_2$CH$_2$ (e), CHCH (f), SiH$_3$CH$_3$ (g), SiH$_2$CH$_2$ (h), and SiHCH (i). The color of the bonds indicates the mutual information of the connected two sites. The shape of the important molecular orbitals (isovalue 0.05) are shown on the two sides where numbers near the orbitals indicate the site number. The blue and red colors of the orbitals refer to the sign of the wave function. Atoms are labeled with their usual colors; H, B, C, N, and Si atoms marked with white, brown, tan, blue, and yellow colors, respectively.
}
\label{fig:series}
\end{figure*}

{\sl Multiple bonds:}
We investigate N$_2$ and CO as model compounds for multiple bond systems (Fig.~\ref{fig:small}g, h).
Indeed, the mutual information analysis results in three bonds for both
molecules as expected but detailed investigation identify fundamental
differences. For $\pi$-bond in N$_2$, the eigenvector elements,
$c^{(5,8)}_1(2,0)=c^{(3,10)}_1(2,0)=0.573\times[0.72, 1,  -1,  0.72]$,
are similar to the previously discussed covalent bonds with the dominance of electron-sharing components.
While the $\pi$ bonds of CO, however, have strong asymmetric distribution
in eigenvector elements,
$c^{(5,8)}_1(2,0)=c^{(4,9)}_1(2,0)=0.644\times[0.36,  0.80, -0.80, 1]$, which surpass
the usual asymmetry resulted from the polarization of the bond.
Therefore, we conclude that QIT analysis may be able to differentiate
between covalent and donor-acceptor bonds as well.

{\sl Dative and multiple bonds:}
Unfortunately, owing to the symmetry of the CO molecule we cannot examine
separately the two $\pi$ bonds (one covalent and one donor-acceptor
bonds) and support our hypothesis.
To gain deeper insight, we investigate a series of molecules which can help
to elucidate this question. We choose NH$_3$BH$_3$, NH$_2$BH$_2$, and NHBH
for donor-acceptor test systems; CH$_3$CH$_3$, CH$_2$CH$_2$, and CHCH for
isoelectronic apolar covalent reference systems. We also calculate
SiH$_3$CH$_3$, SiH$_2$CH$_2$, and SiHCH as polarized covalent analogs which
also serve as an example for bonds containing heavier element
(see Fig.~\ref{fig:series}). In the rest of the paper, we only focus
on the newly emerged bonding modes, while the previously discussed bond types
are considered as known, if it is not mentioned otherwise. As an example, in this section, all C-H, N-H,
and B-H bonds show similar picture as we have discussed for these type of bonds
in the case of BH$_3$, CH$_4$, and NH$_3$. We note that this result also
indicates that QIT based analysis is robust, yields the same relevant
information for the same chemical moiety. We provide all details
for all investigated molecules in Supporting Information \ref{sec:dative}. 

\begin{figure*}
\centerline{
\includegraphics[scale=1.2]{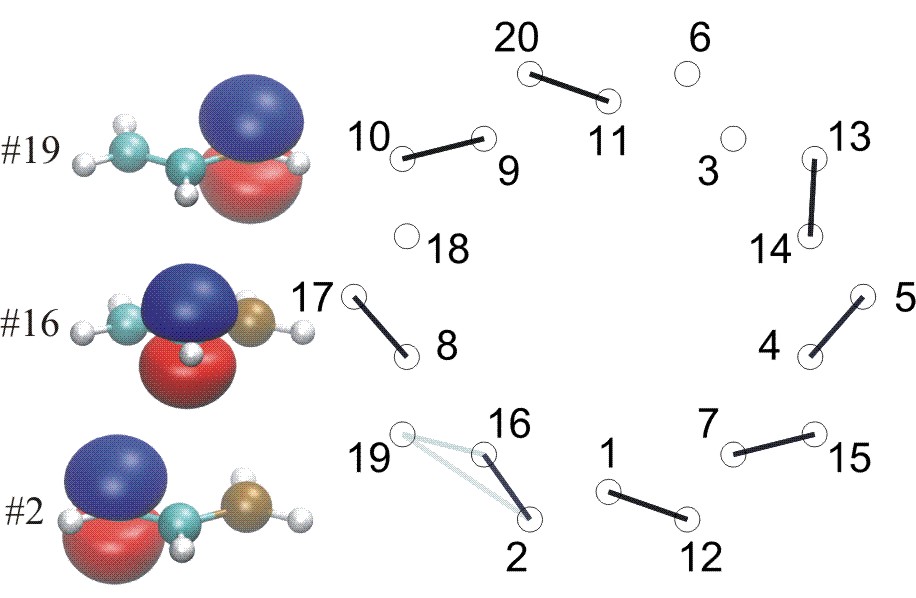}
\hskip 0.8cm
\includegraphics[scale=1.2]{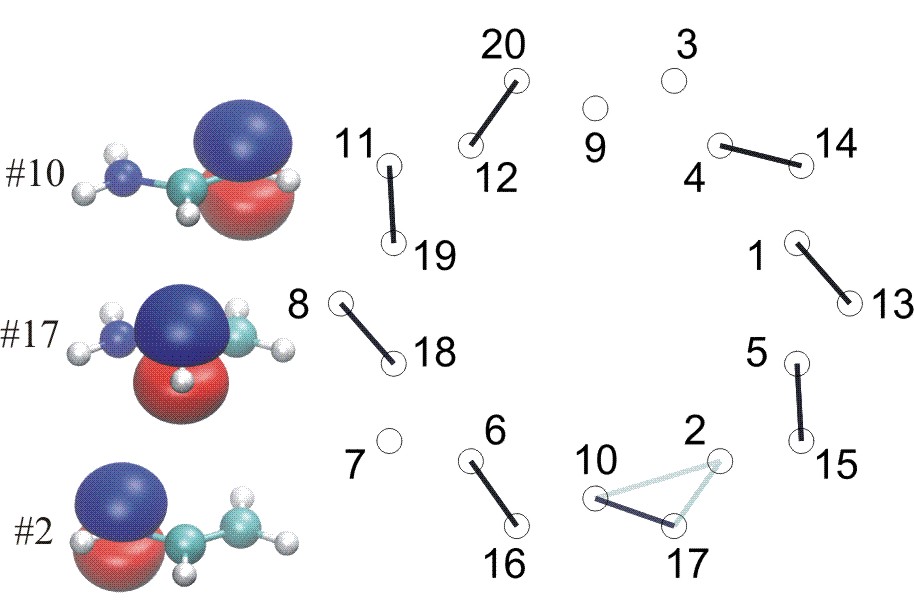}
\hskip 0.8cm
\includegraphics[scale=1.2]{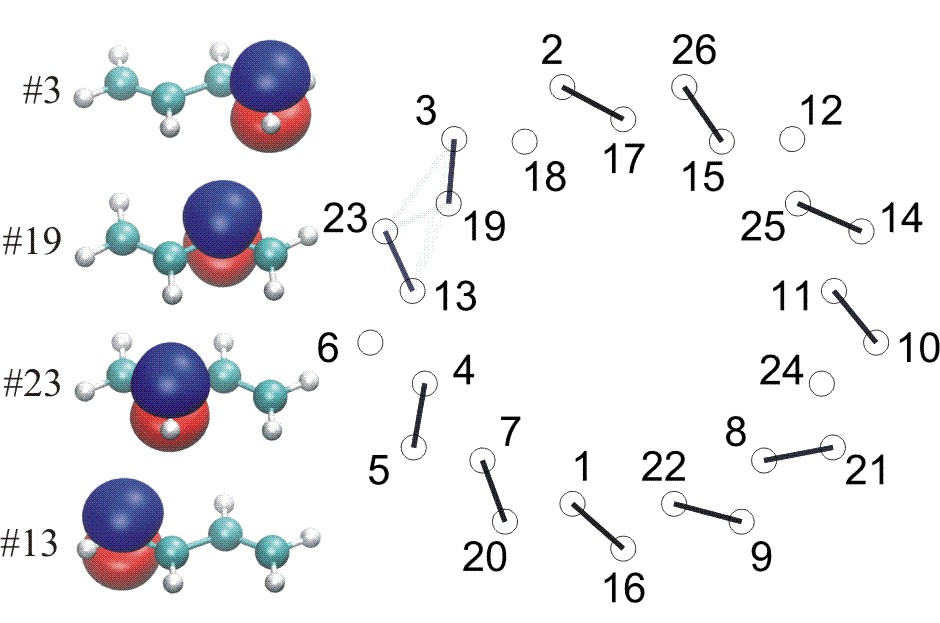}
}
\vskip -3.4cm
\centerline{\hskip -5.4cm a)\hskip 5.5cm b) \hskip 5.2cm c)}
\vskip 3.8cm
\centerline{
\includegraphics[scale=1.2]{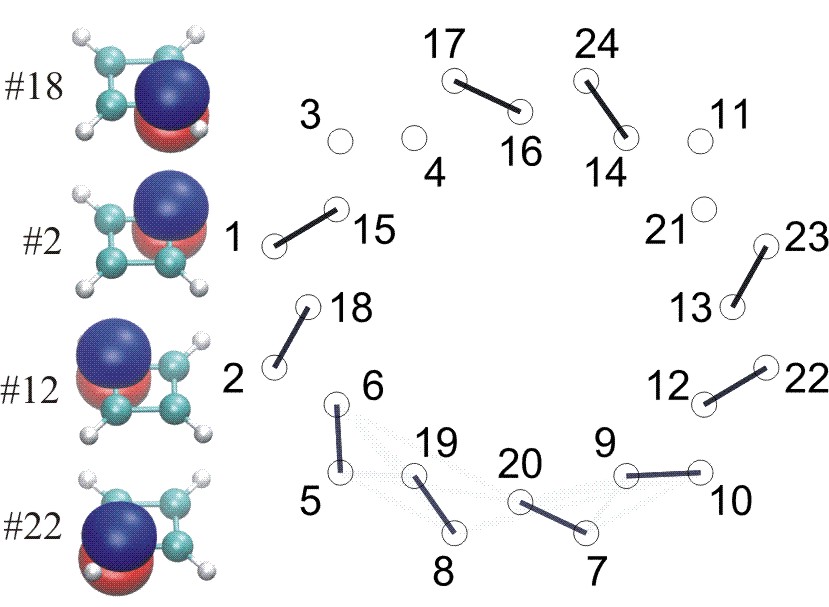}
\hskip 1.8cm
\includegraphics[scale=1.2]{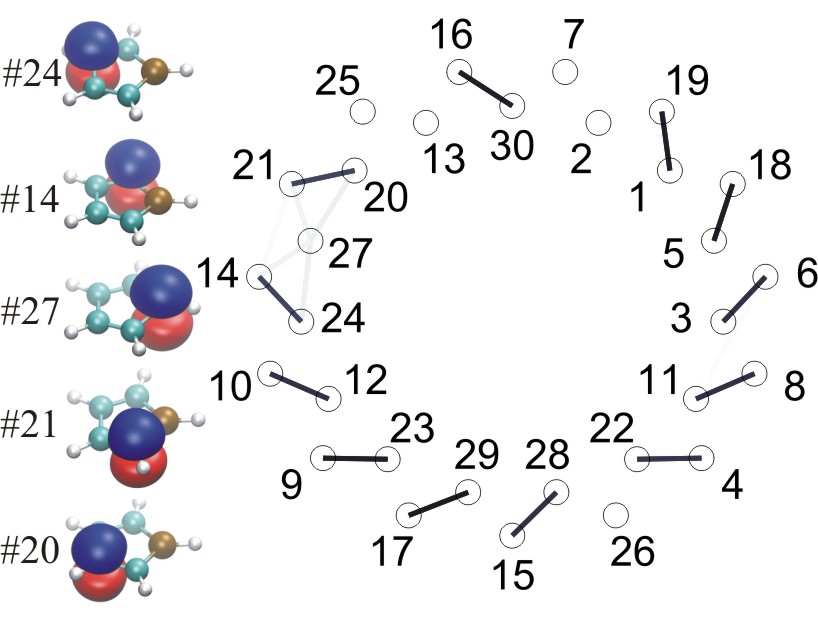}
\hskip 1.4cm
\includegraphics[scale=1.2]{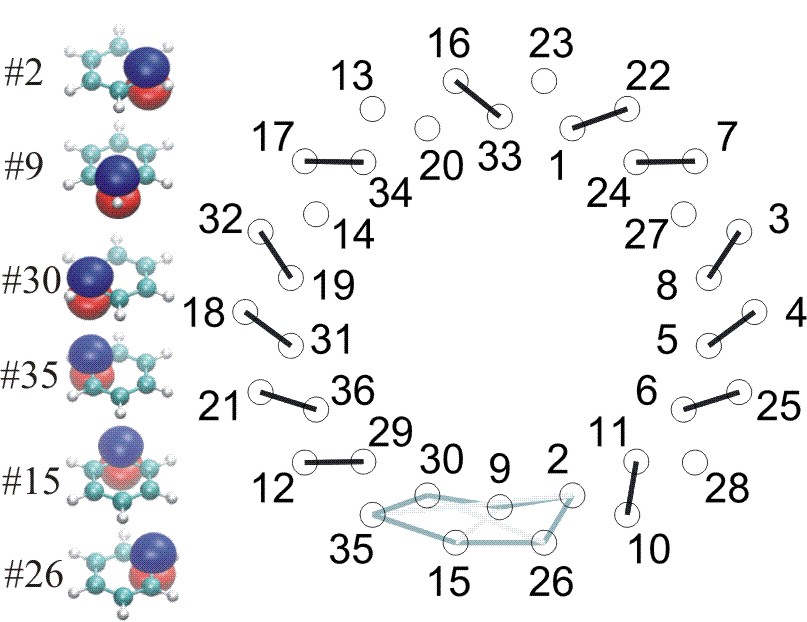}
}
\vskip -3.4cm
\centerline{\hskip -5.4cm d)\hskip 5.5cm e) \hskip 5.2cm f)}
\vskip 3.8cm
\centerline{
\hskip 0.2cm
\includegraphics[scale=1.2]{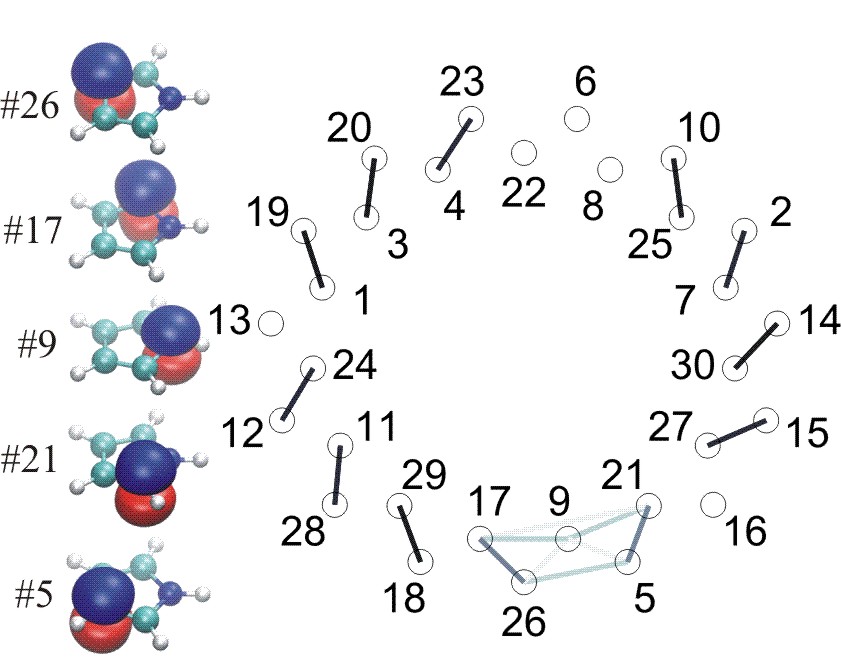}
\hskip 1.8cm
\includegraphics[scale=1.2]{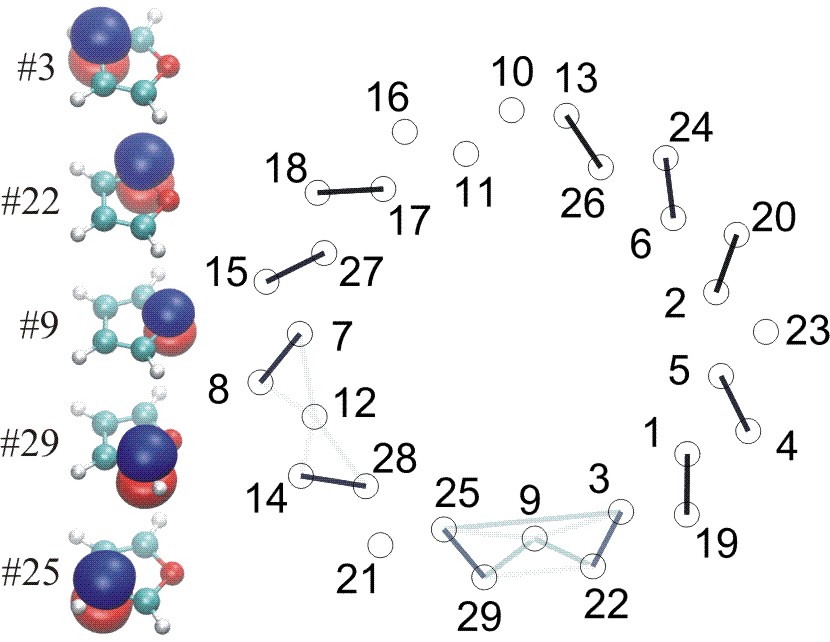}
\hskip 0.6cm
\includegraphics[scale=1.2]{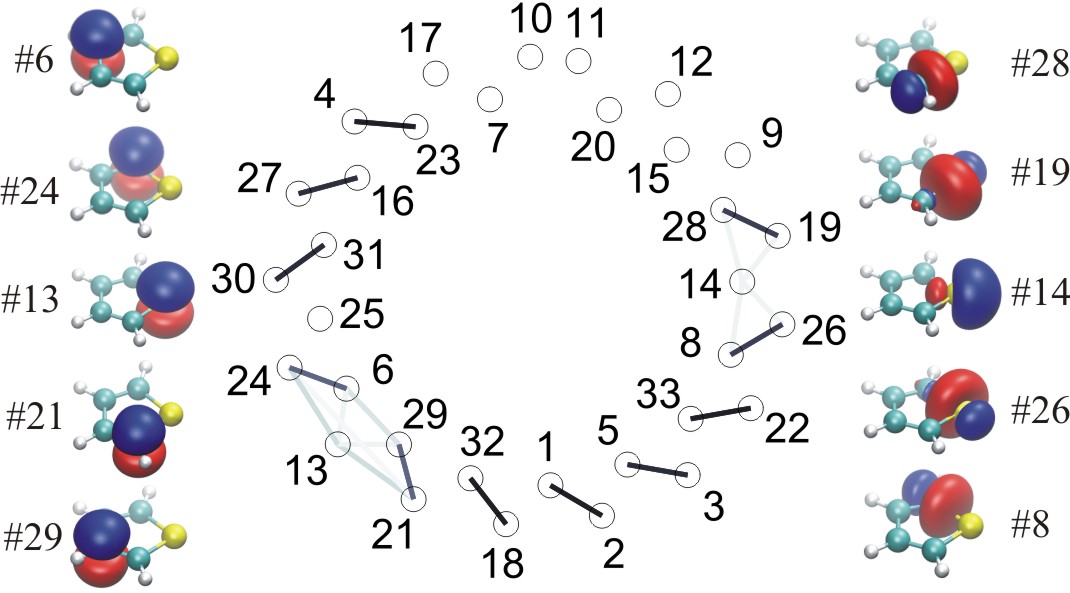}
}
\vskip -3.6cm
\centerline{\hskip -5.4cm g)\hskip 5.5cm h) \hskip 5.2cm i)}
\vskip 4.0cm
\caption{Graphical representation of mutual information for the series of delocalized systems: CH$_2$CHBH$_2$ (a), CH$_2$CHNH$_2$ (b), butadiene (c), cyclobutadiene (d), borole (e), benzene (f), pyrrole (g), furan (h), and thiophene (i). The color of the bonds indicates the mutual information of the connected two sites. The shape of the important molecular orbitals (isovalue 0.05) are shown on the two sides where numbers near the orbitals indicate the site number. The blue and red colors of the orbitals refer to the sign of the wave function. Atoms are labeled with their usual colors; H, B, C, N, O, and S atoms marked with white, brown, tan, blue, red, and yellow colors, respectively. 
}
\label{fig:deloc}
\end{figure*}

QIT analysis for NH$_3$BH$_3$ (Fig.~\ref{fig:series}a)
provides the similar results to the donor-acceptor bond of CO.
The mutual information of the donor-acceptor bond is
significantly lower than for the covalent bond,
$I^{(5,13)}=1.75$.
The eigenvector elements show drastic asymmetric distribution and 
the weight of electron-sharing parts is also smaller than in the previous examples:
$c^{(5,13)}_1(2,0)=0.762\times[0.23,   -0.58,    0.58,    1]$. For the isoelectronic
analog CH$_3$CH$_3$ (Fig.~\ref{fig:series}d), the C-C bond, site \#5 and \#13, has similar values
to other covalent bonds; $I^{(5,13)}$ = 2.49, the eigenvector elements indicate
difference from donor-acceptor bonding mode $c^{(5,13)}_1(2,0)=0.542\times [0.84,  1,  -1,  0.84]$. 
SiH$_3$CH$_3$ with polarized Si-C covalent bond shows the expected results (Fig.~\ref{fig:series}g);
mutual information is larger ($I^{(17,18)}=2.28$) than that of the
donor-acceptor analog but somewhat smaller than that of the homonuclear
analog. The important eigenvector elements
show the expected asymmetry 
$c^{(17,18)}_1(2,0)=0.553\times [1,  -0.96, 0.96,  0.64]$.
Such direct comparison can be flawed by the large electronegativity 
difference of atoms
therefore we investigate NH$_2$BH$_2$. 
The mutual information analysis suggests two strong
interactions between N and B atoms (Fig.~\ref{fig:series}b),
$I^{(3,4)}$ and $I^{(6,11)}$,
in accordance with the double bond structure. Interestingly,
the QIT results of these bonds are quite different.
On one hand, $I^{(3,4)}$ is 2.38 close to the values of
previously mentioned covalent bonds and the eigenvector elements of the
largest $\omega^{(3,4)}$ also supports the polarized covalent bond
assignation based on previous examples
$c^{(3,4)}_1(2,0)=0.628\times[1,  0.80, -0.80,  0.51]$.
On the other hand,
$I^{(6,11)}=1.94$ is much lower and the eigenvector elements
is determined by the strongly asymmetric distribution
$c^{(6,11)}_1(2,0)=0.768\times[0.16, 0.58,  -0.58,  1]$ similar to the
donor-acceptor dative bond in NH$_3$BH$_3$. Assignation based on orbital
images proves that the aforementioned results are consistent with the
chemical picture; the $\pi$ bond is the dative bond while the $\sigma$ is
the covalent bond in NH$_2$BH$_2$. Although, these results indicate no clear cut between covalent
and dative bonds but the difference is prominent which is enough to assign them even within a double bond. 
Analog molecules CH$_2$CH$_2$ (Fig.~\ref{fig:series}e) and SiH$_2$CH$_2$ (Fig.~\ref{fig:series}h)
show the expected results indicating covalent bonds;
$I^{(i,j)}$ for these four bonds is in the range of 2.3-2.5
(see details in Supporting Information \ref{subsec:ch2ch2}, \ref{subsec:sih2ch2}).
For NHBH (Fig.~\ref{fig:series}c), a sigma bond
is given by $I^{(2,3)}$ = 2.48,
$c^{(2,3)}_1(2,0)=0.602\times[1, 0.84, -0.84,  0.60]$, while the two
$\pi$ bonds are identical, because of symmetry reasons
($I^{(5,9)}=I^{(6,10)} = 1.91$,
$c^{(5,9)}_1(2,0)=c^{(6,10)}_1(2,0)=0.644\times [0.35,   0.80, -0.80, 1]$).
These results
are nearly the same as the results of CO 
suggesting the same bonding mode.
For analog CHCH (Fig.~\ref{fig:series}f), we obtain almost the same results as for N$_2$;
for the $\sigma$ bond: $I^{(2,3)}$ = 2.61, $c^{(2,3)}_1(2,0)=0.516\times [0.94,   1, -1, 0.94]$,
and for the $\pi$ bonds: $I^{(5,10)}$ = $I^{(6,9)}$ =2.03, $c^{(5,10)}_1(2,0)=c^{(6,9)}_1(2,0)=0.571\times [0.73,  -1, 1, 0.73]$.
The same is valid for the analysis of SiHCH (Fig.~\ref{fig:series}i) with the expected slight
polarization (see in SI \ref{subsec:sihch}).

{\sl Delocalized systems:}
After the successful determination of multiple and dative bonding modes,
we extend our investigation toward delocalized systems. As a transition,
we continue with the examination of CH$_2$CH$_2$BH$_2$ and
CH$_2$CH$_2$NH$_2$ molecules (Fig.~\ref{fig:deloc}a and \ref{fig:deloc}b,
respectively).
Mutual information analysis of CH$_2$CH$_2$BH$_2$ reveals a cyclic structure
between site \#2, \#16, and \#19. There is strong entanglement between
site \#2 and \#16, $I^{(2,16)} = 2.06$,
$c^{(2,16)}_1(2,0)=0.594\times[0.69, -1, 1, 0.61]$,
which corresponds to the C-C $\pi$ bond. While the other two mutual information data
are an order of magnitude smaller, $I^{(2,19)} = 0.37$, $I^{(16,19)} = 0.43$.
Further analysis points out that the main component arise from the one
electron regime: $c^{(16,19)}_1(1,
\pm\frac{1}{2})=0.947\times[-1, 0.34]$,
$c^{(2,19)}_1(1,\pm\frac{1}{2})=0.971\times[1,  0.25]$.
This is in accordance with the
conjugated picture; the vacant orbital of the boron interacts with the
$p_\pi$-orbital carbon atoms and forms a 3-center, 2-electron bond.
Similar effects are found for CH$_2$CH$_2$NH$_2$ but the secondary
interaction stem from the three electron regime:
$c^{(2,17)}_1(3,\pm\frac{1}{2})=0.934\times[ -1, 0.38]$ (see details in SI \ref{subsec:c2h4nh2})
as expected for a 3-center, 4-electron bond.

To investigate longer delocalized systems, we choose butadiene (Fig.~\ref{fig:deloc}c).
There are four interconnected sites in the mutual
information picture. Two strong correlations, $I^{(3,19)} = I^{(13,23)} = 2.13$,
are associated with the C-C $\pi$ bonds and there are secondary
effects between them given by $I^{(19,23)} = 0.16$ 
and $I^{(3,13)} = 0.12$ which is consistent with the
chemical picture of the delocalized $\pi$-system. 
Analyzing longer delocalized molecule like hexatriene shows similar results;
there are secondary effects between the neighboring strong C-C $\pi$ bonds
(see in SI \ref{subsec:hexatriene}). 

{\sl Aromaticity:}
Molecules with cyclic delocalization have special place in chemists’ mind
as the subject of the concept of aromaticity and 
antiaromaticity.
Therefore, we extended our investigation to this direction. From butadiene
and hexatriene one can derive the prototype antiaromatic and aromatic systems,
cyclobutadiene and benzene (Fig.~\ref{fig:deloc}c and \ref{fig:deloc}e,
respectively, and in 
SI \ref{sec:antiaromaticity} and \ref{sec:aromaticity}). For
cyclobutadiene, mutual information analysis indicates two strong correlations
in the $\pi$-system, $I^{(2,18)} = I^{(12,22)} = 2.33$. Interestingly,
the secondary effects observed in butadiene is disappeared in
cyclobutadiene. We have not found any communication between the
$\pi$-subsystems within numerical accuracy, only the $\sigma$-system show some 
minor effects probably due to their strained structure. Another antiaromatic compound
borole shows the similar effects (Fig.~\ref{fig:deloc}e).
In the $\pi$-system two C-C-B moieties are found, similar to CH$_2$CH$_2$BH$_2$.
However, there is no secondary effect in the butadiene moiety. Opposite effects 
can be seen for benzene. The strong mutual information between C-C $\pi$-bonds become less dominant
while secondary effects are even stronger compared to hexatriene
and thus a cyclic structure is seen in the mutual information picture 
with very low value
$I^{(2,9)} = I^{(9,30)} = I^{(30,35)} = I^{(35,15)} = I^{(15,26)} = I^{(26,2)} = 0.892 \pm 0.007$,
(Fig.~\ref{fig:deloc}f). Interestingly, other weak interactions are
found for the
opposite sites, at para position of the aromatic ring,
$I^{(2,35)} = I^{(9,15)} = I^{(26,30)} = 0.18$, indicating their direct
relationship. We have also investigated the aromatic series of furan,
pyrrole, and thiophene (Fig.~\ref{fig:deloc}g-i). Interestingly, five-membered 
cyclic structure dominates the mutual information picture of the pyrrole $\pi$-system which is in wide contrast to borole. 
The secondary effects increased between butadiene moiety while the CH$_2$CH$_2$NH$_2$ structure is less emphasized.
Cyclic delocalization with reduced
mutual information between C-C double bonds and increased secondary
correlations are found for furan and thiophene as well indicating similar effects to benzene and pyrrol.
In the mutual information picture of furan and thiophene an additional weak structure is also found which
are the hyperconjugative interaction of the lone pair with the adjacent
$\sigma$-bonds similar to what has been found for water.  
%


In conclusion, we have introduced a novel approach 
to extract information from molecules based on QIT analysis. Systematic
investigation of handful of molecules using localized orbitals offers the
introduction of the concept of chemical bond and aromaticity. We have
shown on several examples how different chemical models like covalent bond,
donor-acceptor dative bond, multiple bond, charge-shift bond, conjugation, and aromaticity follows
from QIT. The discussed results indicate the unified picture of chemical
concepts and therefore can help to elucidate their fundamental features
and may lead to an improved definition of chemical bond.
This study also closes the gap between state-of-the-art physical and
traditional chemical models showing their mutual origin.
Although there are many open questions regarding the QIT analysis of molecules
we envision that our approach can be used in the future  
alternatively or together
with well known chemical bond analysis 
methods\cite{Mayer-1983,Bader-1994,Hiberty-2008,Landis-2012,Frenking-2012}
and elucidate unique bonding modes.~\cite{Hiberty-2012,Hiberty-2013}\\
\\
{\bf Methods}\\
\\
In this paper we have performed numerical calculations using
the quantum chemistry version of the
density matrix renormalization group (QC-DMRG)
method.~\cite{White-1992,White-1999}
We have controlled the numerical accuracy using the 
dynamic block state selection (DBSS) 
approach \cite{Legeza-2003a} and the maximum
number of block states varied in the range of 500-2000 for an
a priory set quantum information loss threshold value $\chi=10^{-5}$.
The ordering of molecular orbitals along the one-dimensional topology of
the DMRG was optimized using the Fiedler approach
\cite{Barcza-2011,Fertitta-2014} and the active space was
extended dynamically based on the dynamically extended active space (DEAS)
procedure.~\cite{Legeza-2003c}

Geometries have been optimized at HF/STO-3G level of theory
which yielded sufficient geometries in accordance with higher level methods.
In QIT, we are especially interested in the
communication of subsystems which may communicate through active sites belonging to
different subsystems.
In molecules, atoms seem a natural choice for the definition of subsystem. 
To create sites which correspond to one atom, we have
applied localized orbitals as sites. We have chosen 
Pipek-Mezey procedure\cite{Pipek-1989}
with tight threshold $10^{-12}$ and minimized the number
of atomic orbitals contributed in each localized orbitals.
Therefore, they can be easily
identified as part of a subsystem and because of the minimal basis set
their chemical meaning is also clear which we can use for later comparison.
All localized orbitals have been used in the DMRG procedure
thus, as a result, we have carried out calculations at the FCI limit for all molecules.
Then, results at the FCI limit have been analyzed in the paper.
All preliminary calculations have been done by MOLPRO Version 2010.1\cite{MOLPRO}.

\begin{acknowledgements}
We thank Sz.\ Szalay and L. Veis for useful discussions.
G.B. and \"O.L. were supported in part by 
the Hungarian Research Fund (OTKA) through Grant
Nos.~K100908 and ~NN110360.
T.S. acknowledge support from the The New Sz\'echenyi Plan TAMOP-4.2.2/B-10/1-2010-0009\\
\end{acknowledgements}

\clearpage
\widetext
\newpage
\setcounter{page}{1}
\makeatletter
\begin{center}
\textbf{\LARGE Supplemental Materials: Concept of chemical bond and aromaticity based on quantum information theory}
\vskip 1cm
\textbf{\large  T. Szilv\'asi, G. Barcza, \"O. Legeza}
\end{center}
{\tiny

\setprefix{f2}{F_2}
\setprefix{n2}{N_2}
\setprefix{co}{CO}
\setprefix{hf}{HF}
\setprefix{lif}{LiF}
\setprefix{naf}{NaF}
\setprefix{co2}{CO_2}
\setprefix{h2o}{H_2O}
\setprefix{bh3}{BH_3}
\setprefix{nh3}{NH_3}
\setprefix{ch4}{CH_4}
\setprefix{cf4}{CF_4}

\setprefix{chch}{CHCH}
\setprefix{ch2ch2}{CH_2CH_2}
\setprefix{ch3ch3}{CH_3CH_3}
\setprefix{sihsih}{SiHSiH}
\setprefix{sih2sih2}{SiH_2SiH_2}
\setprefix{sih3sih3}{SiH_3SiH_3}
\setprefix{sihch}{SiHCH}
\setprefix{sih2ch2}{SiH_2CH_2}
\setprefix{sih3ch3}{SiH_3CH_3}
\setprefix{nhbh}{NHBH}
\setprefix{nh2bh2}{NH_2BH_2}
\setprefix{nh3bh3}{NH_3BH_3}
\setprefix{f2}{F_2}

\setprefix{c2h4bh2}{CH_2CH_2BH_2}
\setprefix{c2h4nh2}{CH_2CH_2NH_2}
\setprefix{butadiene}{Butadiene}
\setprefix{cyclopentadiene}{Cyclopentadiene}
\setprefix{hexatriene}{Hexatriene}

\setprefix{benzene}{Benzene}
\setprefix{pirrol}{Pyrrole}
\setprefix{furane}{Furan}
\setprefix{thiophene}{Thiophene}

\setprefix{cyclobutadiene}{Cyclobutadiene}
\setprefix{borole}{Borole}

\title{Concept of chemical bond and aromaticity based on quantum information theory}
\date{}
\author{T. Szilv\'asi, G. Barcza, \"O. Legeza \vspace{1.0cm}\\ \Large {\bf Supporting Information}}
\maketitle
\let\cleardoublepage\clearpage
\def\triv{bh3, ch4, nh3, h2o, hf, f2, n2, co}
\def\dativ{nh3bh3,nh2bh2,nhbh,ch3ch3,ch2ch2,chch,sih3ch3,sih2ch2,sihch}
\def\deloc{c2h4bh2,c2h4nh2,butadiene,cyclopentadiene,hexatriene}
\def\arom{benzene,pirrol,   furane, thiophene}
\def\antiarom{cyclobutadiene, borole}

\large{
In the Supporting Information (SI) we present the graphical representation
of the single-site entropy $S(\rho^{(i)})$ and the correspondig 
$\omega_\alpha^{(i)}$ eigenvalue spectrum ($\alpha=1\ldots 4$)
as a function of orbital index ($i$).
Furthermore, we show pair-wise elements of the two-orbital 
mutual information ($I^{(i,j)}$) indicated by the
color of the lines between the sites. The lack of strong visible line
between sites is the sign of negligible communication; these orbital
correlations do not contain relevant information. The maximum of the
mutual information is marked by the theoretical limit, $\ln 16\simeq 2.77$.
The largest $\omega^{(i,j)}_\alpha>0.1$ eigenvalues of the corresponding 
two-orbital reduced matrix $\rho^{(i,j)}$ are collected in a table format
together with the eigenvector coefficients obtained in 
the different quantum number sectors of the two-site subsystem.
}

\section{Molecules with covalent bonds}
\label{sec:covalent}
\foreach  \n in \triv{
\begin{figure}[H]
\subsection{${\rm \getprefix{\n}}$}
\label{subsec:\n}
\hskip -1cm
\begin{minipage}[!H]{1.2\textwidth}
\begin{minipage}[t]{0.4\textwidth}
\begin{center}
\includegraphics[scale=0.35]{\n/\n_sto3g_locao_S1complex.png}
\vskip 1cm
\includegraphics[scale=0.25]{\n/\n_sto3g_locao.png}\\
\end{center}
\end{minipage}
\hskip 1cm
\begin{minipage}[t]{0.55\textwidth}
\input{\n/\n_sto3g_locao_bonddata_new_limitomega0.1_nocc_footnotesize.txt}
\end{minipage}
\end{minipage}
\end{figure}
}

\section{Dative systems}
\label{sec:dative}
\foreach  \n in \dativ{
\begin{figure}[H]
\subsection{${\rm \getprefix{\n}}$}
\label{subsec:\n}
\hskip -1cm
\begin{minipage}[!H]{1.2\textwidth}
\begin{minipage}[t]{0.4\textwidth}
\begin{center}
\includegraphics[scale=0.35]{\n/\n_sto3g_locao_S1complex.png}
\vskip 1cm
\includegraphics[scale=0.25]{\n/\n_sto3g_locao.png}\\
\end{center}
\end{minipage}
\hskip 1cm
\begin{minipage}[t]{0.55\textwidth}
\input{\n/\n_sto3g_locao_bonddata_new_limitomega0.1_nocc_footnotesize.txt}
\end{minipage}
\end{minipage}
\end{figure}
}

\section{Delocalized systems}
\label{sec:delocalized}
\foreach  \n in \deloc{
\begin{figure}[H]
\subsection{${\rm \getprefix{\n}}$}
\label{subsec:\n}
\hskip -1cm
\begin{minipage}[!H]{1.2\textwidth}
\begin{minipage}[t]{0.4\textwidth}
\begin{center}
\includegraphics[scale=0.35]{\n/\n_sto3g_locao_S1complex.png}
\vskip 1cm
\includegraphics[scale=0.25]{\n/\n_sto3g_locao.png}\\
\end{center}
\end{minipage}
\hskip 1cm
\begin{minipage}[t]{0.55\textwidth}
\input{\n/\n_sto3g_locao_bonddata_new_limitomega0.1_nocc_footnotesize.txt}
\end{minipage}
\end{minipage}
\end{figure}
}

\section{Antiromaticity}
\label{sec:antiaromaticity}
\foreach  \n in \antiarom{
\begin{figure}[H]
\subsection{${\rm \getprefix{\n}}$}
\label{subsec:\n}
\hskip -1cm
\begin{minipage}[!H]{1.2\textwidth}
\begin{minipage}[t]{0.4\textwidth}
\begin{center}
\includegraphics[scale=0.35]{\n/\n_sto3g_locao_S1complex.png}
\vskip 1cm
\includegraphics[scale=0.25]{\n/\n_sto3g_locao.png}\\
\end{center}
\end{minipage}
\hskip 1cm
\begin{minipage}[t]{0.55\textwidth}
\input{\n/\n_sto3g_locao_bonddata_new_limitomega0.1_nocc_footnotesize.txt}
\end{minipage}
\end{minipage}
\end{figure}
}

\section{Aromaticity}
\label{sec:aromaticity}
\foreach  \n in \arom{
\begin{figure}[H]
\subsection{${\rm \getprefix{\n}}$}
\label{subsec:\n}
\hskip -1cm
\begin{minipage}[!H]{1.2\textwidth}
\begin{minipage}[t]{0.4\textwidth}
\begin{center}
\includegraphics[scale=0.35]{\n/\n_sto3g_locao_S1complex.png}
\vskip 1cm
\includegraphics[scale=0.25]{\n/\n_sto3g_locao.png}\\
\end{center}
\end{minipage}
\hskip 1cm
\begin{minipage}[t]{0.55\textwidth}
\input{\n/\n_sto3g_locao_bonddata_new_limitomega0.1_nocc_footnotesize.txt}
\end{minipage}
\end{minipage}
\end{figure}
}

}

\end{document}